\documentclass[preprint,12pt]{elsarticle}
\usepackage{graphicx}
\usepackage{float}
\usepackage[ruled,linesnumbered]{algorithm2e}
\usepackage{amssymb}
\usepackage{amsmath}
\usepackage{natbib}
\usepackage{lipsum}
\UseRawInputEncoding
\journal{Journal of Computational Science}

\begin{document}
\begin{frontmatter}

\title{PHWSOA: A Pareto-based Hybrid Whale-Seagull Scheduling for Multi-Objective Tasks in Cloud Computing}

\author[a]{Zhi Zhao} \ead{zhaozh32@163.com} 
\author[a]{Hang Xiao} \ead{nirxiaohang@163.com}
\author[a]{Wei Rang \corref{mycorrespondingauthor}} \ead{wrang@sdnu.edu.cn}
%% Author affiliation
\cortext[mycorrespondingauthor]{Corresponding author}
\address[a]{School of Information Science and Engineering, Shandong Normal University, Jinan, 250000, China}

%% Abstract
\begin{abstract}
Task scheduling is a critical research challenge in cloud computing, a transformative technology widely adopted across industries. Although numerous scheduling solutions exist, they predominantly optimize singular or limited metrics such as execution time or resource utilization often neglecting the need for comprehensive multi-objective optimization. To bridge this gap, this paper proposes the Pareto-based Hybrid Whale-Seagull Optimization Algorithm (PHWSOA). This algorithm synergistically combines the strengths of the Whale Optimization Algorithm (WOA) and the Seagull Optimization Algorithm (SOA), specifically mitigating WOA's limitations in local exploitation and SOA's constraints in global exploration. Leveraging Pareto dominance principles, PHWSOA simultaneously optimizes three key objectives: makespan, virtual machine (VM) load balancing, and economic cost. Key enhancements include: Halton sequence initialization for superior population diversity, a Pareto-guided mutation mechanism to avert premature convergence, and parallel processing for accelerated convergence. Furthermore, a dynamic VM load redistribution mechanism is integrated to improve load balancing during task execution. Extensive experiments conducted on the CloudSim simulator, utilizing real-world workload traces from NASA-iPSC and HPC2N, demonstrate that PHWSOA delivers substantial performance gains. Specifically, it achieves up to a 72.1\% reduction in makespan, a 36.8\% improvement in VM load balancing, and 23.5\% cost savings. These results substantially outperform baseline methods including WOA, GA, PEWOA, and GCWOA underscoring PHWSOA's strong potential for enabling efficient resource management in practical cloud environments.
\end{abstract}


\end{frontmatter}

\section{Introduction}
Cloud computing offers on-demand access to virtualized resources, including computing power, storage, and network bandwidth, enabling the deployment of flexible and scalable IT infrastructures \cite{1}. In this context, efficient task scheduling is critical for optimal resource allocation and the fulfillment of key performance objectives. The scheduling process entails mapping computational tasks to appropriate VM while meeting Quality-of-Service(QoS) constraints and optimizing metrics such as makespan and operational cost \cite{2}. However, persistent challenges complicate this process. Task heterogeneity and the diverse capabilities of VM make resource allocation decisions inherently complex \cite{3}. Furthermore, simultaneously optimizing conflicting objectives like minimizing execution time and cost poses a significant challenge, requiring the identification of trade-offs among Pareto-optimal solutions \cite{4}. Consequently, developing intelligent scheduling strategies capable of navigating this complex multi-objective landscape remains a vital research challenge in cloud computing.

To address these challenges, this paper proposes a novel hybrid metaheuristic framework named the PHWSOA. Our main contributions are summarized as follows:

(1) We propose PHWSOA, a novel hybrid framework that synergistically integrates the global exploration capabilities of the WOA with the local exploitation strengths of the SOA, specifically designed to address complex multi-objective task scheduling problems in cloud environments.

(2) Within this framework, we develop enhanced mechanisms including Halton sequence initialization for superior population diversity, an adaptive VM load redistribution strategy for dynamic resource management, parallel processing for accelerated convergence, and a Pareto-guided mutation operator to escape local optima, collectively improving scheduling efficiency and solution quality.

(3) We design the algorithm to construct a comprehensive Pareto-optimal solution set through dominance-based selection, enabling the effective co-optimization of three conflicting objectives: makespan minimization, execution cost reduction, and load balance improvement.

(4) Through extensive simulations using the CloudSim toolkit with real-world workload traces (NASA-iPSC and HPC2N), we demonstrate that PHWSOA significantly outperforms established baselines (WOA, GA, PEWOA, GCWOA), achieving notable improvements including up to a 72.1\% reduction in makespan, a 36.8\% enhancement in load balance efficiency, and a 23.5\% decrease in execution cost.

The remainder of the paper is organized as follows. Section 2 reviews related work on cloud task scheduling. Section 3 presents the system model and problem formulation. Section 4 details the mathematical foundations of the base algorithms (WOA and SOA). Section 5 elaborates on the proposed PHWSOA framework and its key enhancements. Section 6 discusses the experimental setup, results, and comparative analysis. Finally, Section 7 concludes the paper and suggests directions for future research.

\section{Related work}
The increasing adoption of cloud computing across industries has intensified the demand for systems that deliver superior performance, enhanced QoS, and greater energy efficiency. To meet these complex requirements, a variety of multi-objective optimization algorithms have been proposed to enhance the operational efficiency of cloud platforms.

Notable recent contributions include the following. Anup Gade et al. \cite{5} introduced a scheduling method that integrates task clustering with parameter normalization and a Normalized Adaptive League Championship Algorithm (NALCA) to minimize makespan. Their approach utilizes hierarchical clustering based on equal-probability modeling for efficient task grouping. Xinqi Qin et al. \cite{6} developed an Enhanced Red-tailed Hawk Algorithm (ERTH), which incorporates multi-elite selection and chaotic mapping techniques. Their method achieved significant performance gains, reducing overall system costs by 34.8-36.4\% compared to traditional schedulers in a custom simulation environment.

Within the specific domain of workload balancing, several advanced techniques have been developed. Santosh Kumar Paul et al. \cite{7} proposed an Improved Artificial Rabbit Optimization with Pattern Search (IARO-PS) approach, which combines task partitioning with optimal mapping strategies to enhance resource efficiency. Similarly, Shaimaa Badr et al. \cite{8} introduced a Task-Combining Power Minimization (TCPM) method that integrates strengths from existing algorithms via a best-fit assignment strategy, demonstrating lower power consumption compared to conventional approaches like FCFS and WWO.

Despite these advancements, prevailing algorithms often exhibit persistent limitations such as slow convergence rates and a propensity for premature convergence to local optima. These shortcomings typically lead to compromised performance on critical metrics, including makespan, latency, and computational overhead. Consequently, a clear need exists for novel hybrid meta-heuristic frameworks capable of effectively balancing multiple, often conflicting, optimization objectives without relying on prior knowledge of task or resource characteristics. To bridge this gap, our study introduces the PHWSOA. This framework incorporates an adaptive load-balancing mechanism and is specifically designed to minimize makespan while enhancing overall system performance through a refined balance between global exploration and local exploitation.

\section{System Architecture and Scheduling Issues}
\subsection{System design}
Task scheduling constitutes a fundamental component of cloud computing infrastructure, responsible for efficiently mapping user-submitted computational tasks to appropriate virtualized resources. As illustrated in Figure 1, the scheduling process can be abstracted into five critical stages:

(1) Workflow Initiation: The process begins with a set of tasks awaiting execution, each defined by its computational profile and constraints.

(2) Resource Estimation and Allocation: The system evaluates each task's computational demands while monitoring the status of cloud resources, including VM availability, host load, and network bandwidth. Based on this assessment, it dynamically allocates VMs to suitable physical hosts \cite{9}.

(3) Scheduling Decision: An optimization algorithm assigns tasks to specific VMs according to real-time resource demands and VM capacity constraints. This step aims to maximize overall system efficiency while meeting QoS requirements \cite{10}.

(4) Task Execution and Monitoring: During execution, the framework continuously monitors key performance indicators such as progress, resource utilization, and system health. In cases of performance degradation or failure, it triggers recovery mechanisms like live migration or dynamic rescheduling.

(5) Task Completion and Result Delivery: Upon successful completion, results are returned to the user, allocated resources are released back to the pool, and comprehensive execution metrics are logged for subsequent performance analysis and scheduling optimization.

This structured approach ensures efficient resource utilization and maintained service quality across diverse workloads. The complexity inherent in each stage necessitates sophisticated optimization strategies to address the dynamic and uncertain nature of cloud environments.
\begin{figure}[h]
    \centering
    \includegraphics[width=0.7\linewidth]{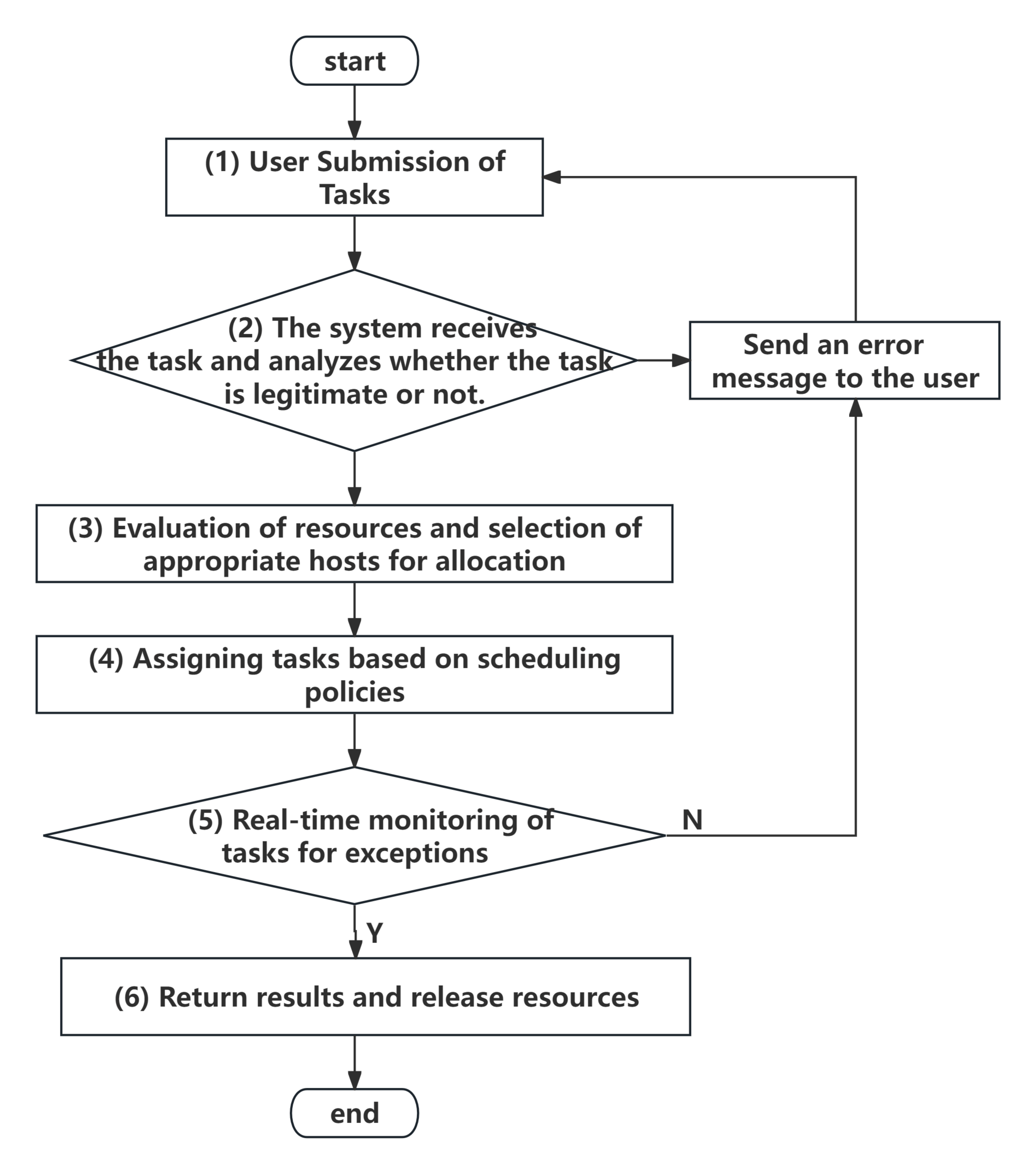}
    \caption{System workflow. }
    \label{fig:enter-labe1}
\end{figure}

\subsection{Problem Statement}
Efficient task scheduling poses a significant challenge in cloud computing due to the dynamic and scalable nature of the infrastructure. Suboptimal strategies often lead to workload concentration on specific VMs, leaving others underutilized. This imbalance results in extended task queuing times, increased response latency, prolonged makespan, and degraded parallel efficiency-collectively compromising overall system performance and QoS.

Moreover, these scheduling inefficiencies exacerbate system-wide load imbalances, where VM overloading coexists with resource idleness. Such conditions not only delay task completion and diminish scheduling effectiveness but also incur unnecessary operational costs under prevalent pay-per-use models. More critically, they increase the risk of violating Service-Level Agreements (SLAs) and substantially degrade the user experience.

In essence, cloud task scheduling constitutes a complex multi-objective optimization problem. It must simultaneously address competing goals-including minimizing execution time, maximizing resource utilization, optimizing load balance, and reducing cost-which frequently involve inherent trade-offs. The development of robust, adaptive, and efficient strategies to navigate these trade-offs therefore remains a crucial research frontier for advancing cloud computing performance.

\section{Algorithm and Modeling}
This section first introduces the foundational mechanisms of the WOA and the SOA. It then provides a formal mathematical formulation of the multi-objective cloud task scheduling problem. Subsequently, the proposed PHWSOA is described in detail, emphasizing its novel optimization strategies and underlying design principles. Finally, the overall workflow and key algorithmic enhancements of PHWSOA are illustrated through a flowchart, elucidating the optimization process and its expected performance benefits.

\subsection{Whale Optimization Algorithm}
The WOA is a bio-inspired metaheuristic that emulates the cooperative foraging behavior of humpback whales. Its search process is governed by three key mechanisms: prey encircling, bubble-net feeding (exploitation), and random exploration.

(1) Prey Encircling: In this phase, whales identify and encircle a target prey. The algorithm assumes the current best candidate solution represents the prey's location, guiding other search agents to converge toward it iteratively. This collective movement is mathematically modeled as follows:
\begin{equation}
    \vec{D}=\begin{vmatrix}\vec{C}\cdot\vec{X^*}(t)-\vec{X}(t)\end{vmatrix}
\end{equation}

\begin{equation}
    \vec{X}(t+1)=\overrightarrow{X^{*}}(t)-\vec{A} \cdot \vec{D}
\end{equation}

Where \textit{t} is the iteration counter, $\vec{X}$ is the whale's position vector, $\vec{X^*}$ is the current best solution, $\vec{D}$ represents the distance to the prey, and $\vec{A}$ and $\vec{C}$ are coefficient vectors guiding the movement direction.

\begin{equation}
\overrightarrow{D^{\prime}}=\left|\overrightarrow{X^*}(t)-\vec{X}(t)\right|
\end{equation}

\begin{equation}
\vec{X}(t+1)=\overrightarrow{D^{^{\prime}}}\cdot e^{bl}\cdot\cos{(2\pi l)}+\overrightarrow{X^{*}}(t)
\end{equation}

where $\overrightarrow{D^{\prime}}$ denotes the Euclidean distance to the prey, \textit{b} defines the spiral's shape, and \textit{l} is a random number in [-1, 1]. 

(3) Randomized Prey Search: In WOA, balancing exploration and exploitation is essential. When the coefficient vector satisfies $\vec{|A|}$ \textgreater 1, the algorithm shifts from local exploitation to global exploration. Whales then update their positions by referencing randomly selected individuals. This behavior is governed by the following equation:
\begin{equation}
    \vec{D_{rand}}=|\vec{C}\cdot\vec{X_{rand}}-\vec{X(t)}|
\end{equation}
\begin{equation}
    \vec{X(t+1)}=\vec{X_{rand}}-\vec{A}\cdot\vec{D_{rand}}
\end{equation}
where $\vec{X_{rand}}$ is the position vector of a randomly selected whale, and $\vec{D_{rand}}$ is its distance to the prey.

\subsection{Seagull Optimization Algorithm}
The SOA is a swarm intelligence metaheuristic that models the collective migratory and aggressive behaviors of seagulls to balance global exploration and local exploitation during optimization.

Migration Behavior: This phase simulates the coordinated long-distance movement of a flock. The SOA achieves this by incorporating three fundamental behavioral rules: (a) avoiding collisions between neighboring agents, (b) aligning movement direction toward the best-performing neighbor (local best), and (c) steering the entire population toward the globally best-known solution. These rules are mathematically formulated as follows:

\begin{equation}
 \vec{C_{s}}=A\cdot\vec{P_{s}(x)}
\end{equation}
where $\vec{C_{s}}$ denotes the adjusted position to avoid neighboring agents $\vec{P_{s}(x)}$, is the current position, and \textit{x} represents the current iteration count \textit{A} regulates the movement dynamics in the search space. After collision avoidance, the agent moves toward the optimal neighbor:
\begin{equation}
 \vec{M_{s}}=\vec{B}\cdot(\vec{P_{bs}(x)}-\vec{P_{s}(x)})
\end{equation}
where $\vec{M_{s}}$ represents the relative position of the search agent $\vec{P_{s}(x)}$ with respect to the optimal search agent $\vec{P_{bs}(x)}$, the parameter \textit{B} regulates an appropriate balance between exploration and exploitation. Finally, the position is updated using information from the globally best agent:
\begin{equation}
 \vec{D_{s}}=|\vec{C_{s}}+\vec{M_{s}}|
\end{equation}
Seagulls dynamically adjust their flight parameters including attack angle, speed, and altitude during migration by modulating wing dynamics and utilizing body mass. When engaging prey, they perform characteristic spiral maneuvers in three-dimensional space. This aggressive attack behavior within a Cartesian (x, y, z) coordinate system is mathematically modeled as follows:
\begin{equation}
x=r\cdot cos(\theta)
\end{equation}
\begin{equation}
y=r\cdot sin(\theta)
\end{equation}
\begin{equation}
z=r\cdot \theta
\end{equation}
\begin{equation}
r=u \cdot e^{\theta v}
\end{equation}
where \textit{r} is the radius of each spiral turn, \textit{k} is a random number uniformly distributed over [0, 2$\pi$], \textit{u} and \textit{v} are spiral shape determining constants, and \textit{e} is the base of the natural logarithm. Based on this model, the updated position of the search agent is computed as:
\begin{equation}
\vec{P_{s}(x)}=(D \cdot x \cdot y \cdot z)+\vec{P_{bs}(x)}
\end{equation}

\subsection{Mathematical modeling}
In cloud computing task scheduling, the mapping between tasks and VMs is encoded as a position vector in PHWSOA, where each element denotes the assignment of task i to VM j. Thus, each position vector defines a specific task-to-VM allocation. In the scheduling model, we consider a population of \(p \) whales managing \(n \) independent tasks and \(m \) VMs, where \(n \gg m \).

\textit{Definition 1: Task}
Each task in cloud computing is assigned a unique identifier ($T_{id}$) for distinguish. Its key attributes include: task length ($T_{length}$), denoting computational demand in million instructions (MI); required number of processing cores ($T_{pesNumber}$); input data size ($T_{fileSize}$); and the output data size ($T_{outSize}$), both measured in bytes (B). A task can be represented as follows: \\$\mathrm{T_{i}=f(T_{id},T_{length},T_{pesNumber},T_{fileSize},T_{outputSize})}$.

\textit{Definition 2: Virtual Machine}

A VM is a fundamental element of cloud resource management, enabling the abstraction of physical resources and facilitating flexible computation and efficient task execution. Each VM is uniquely identified ($VM_{id}$) and characterized by key attributes: computational power ($VM_{mips}$), measured in Million Instructions per Second (MIPS); the number of processing cores($VM_{pesNumber}$). A  can be represented as follows: follows: $$\mathrm{VM_i=f(VM_{id},VM_{mips},VM_{pesNumber})}$$

\textit{Definition 3: Mapping Matrix of Tasks and Virtual Machines}

Each task in Definition 1 is assigned to a single VM, while a VM may handle multiple tasks. All tasks are independent, and the task-to-VM mapping is represented by the following matrix: Let
$X=\begin{bmatrix} x_{11},x_{12},\ldots,x_{1n}\\ x_{21},x_{22},\ldots,x_{2n}\\ \vdots\\ x_{p1},x_{p2},\ldots,x_{pn}\end{bmatrix}$,where X is a \textit{p}*\textit{n} matrix, with \textit{p} representing the population size (set to 50 in this experiment) and \textit{n} representing the number of tasks. Each row $x_{k}=\{x_{k1},x_{k2},...x_{kn}\}$indicates the position of the kth whale, i.e. , a complete mapping scheme from a set of tasks to virtual machines.

\textit{Definition 4: makespan}

The execution time (ET) of task i on VM j is given by the ratio of the task length to the VM's computing power (MIPS). The completion time (CT) for all tasks on VM j is the sum of their lengths divided by the VM's MIPS. The total completion time (makespan) for all tasks is expressed as: $makespan=\max(CT_{j}),1\leq j\leq m$.

\textit{Definition 5: Throughput}
Throughput is defined as the number of tasks completed per unit time and is calculated as: $throughput=\frac{Totalnumber\ of\ tasks}{makespan}$.

\textit{Definition 6: Load}
Load describes the distribution of computational workload across resources (e.g. , virtual machines) and is typically measured by the balance of resource utilization. The total system load is then expressed as: $load=\sum_{i=1}^n\delta(x_{ki},j)\sqrt{\frac{\sum_{j=1}^m(load_j-avgload_j)^2}{m}}$, where$\delta(x_{ki},j)= \begin{cases} 1,\mathrm{ifx}_{ki}=j \\ 0,\mathrm{ifx}_{ki}\neq j & \end{cases}$.

\textit{Definition 7: Cost}
Cost represents the economic overhead associated with task execution, encompassing resource consumption such as computation, storage, and memory. For cloud subscribers, cost is a critical factor in selecting cloud services. Similarly, cloud providers benefit from cost reduction by offering competitive pricing, improving the price-performance ratio, and attracting more users, thereby enhancing market competitiveness. The total cost comprises two components: running cost ($C_{r}$) and bandwidth cost ($C_{bw}$), The running cost $C_{r}$, which includes memory and storage expenses, is expressed by the following formula:
\begin{equation}
    \begin{aligned} C_{\mathrm{r}} & =\sum_{j=1}^{m}(ram\cos t_{j}+storage\cos t_{j})^{*}Time_{j} \\ & =\sum_{i=1}^{m}(ramsize_{j}*P_{1}+storage\cos t_{j}*P_{2})^{*}Time_{j} \end{aligned}
\end{equation} 
Bandwidth cost represents the economic and computational overhead associated with data transmission during task execution. It primarily includes the network resources consumed when transferring data between nodes, along with the corresponding costs. This can be expressed using the following equation: $\begin{aligned} C_{\mathrm{bw}} & =transtime*P_{3} \\ & =\sum_{j=1}^{m}(\frac{Totallength\ of\ VM_{j}}{bw_{j}})^{*}P_{3} \end{aligned}$,
Therefore, the total cost can be expressed as: $\cos t=C_{\mathrm{r}}+C_{\mathrm{bw}}$.

\section{A Hybrid pareto-based whale-seagull optimization algorithm}
While the WOA demonstrates strong global exploration capability, it often lacks precise local refinement, which can lead to slow or premature convergence. Conversely, the SOA provides effective local exploitation but exhibits limited global search potential. To synergistically overcome these complementary limitations, we propose the PHWSOA. This hybrid framework is designed to integrate the exploratory strength of WOA with the exploitative precision of SOA, thereby enhancing overall optimization accuracy and convergence speed.

The proposed PHWSOA incorporates several key enhancements: (1) Halton sequence-based population initialization for improved diversity; (2) a dynamic, VM load-aware resource redistribution strategy; (3) a parallel task execution model to accelerate evaluation; and (4) the integration of genetic mutation operators to preserve population diversity and avert premature convergence. Furthermore, the algorithm employs Pareto dominance principles to construct a non-dominated solution set, enabling the simultaneous optimization of three conflicting objectives: makespan, VM load balancing, and economic cost. Collectively, these mechanisms enhance global search robustness, improve resource efficiency, and accelerate convergence for complex multi-objective cloud scheduling problems.

\subsection{Halton Sequence Population Initialization}
To enhance initial population diversity and global exploration while mitigating premature convergence, we employ the Halton sequence for population initialization instead of conventional random sampling. As a low-discrepancy sequence grounded in number theory, the Halton sequence generates a more uniform distribution of points within the [0,1] interval compared to pseudo-random numbers. This is achieved by converting natural numbers into a base-b numeral system, reversing the resulting digit sequence, and mapping it back to decimal values \cite{11}.

In our implementation, each dimension of the search space is assigned an independent Halton sequence generator, each initialized with a unique prime number base (starting from 2 for the first dimension). The generated points are then linearly scaled to the corresponding dimension's search range [lb, ub]. To ensure distinct initial populations for the hybrid framework, the WOA population is initialized using the first N points from the sequence, whereas the SOA population is initialized using the subsequent N points (starting from index i+N). This systematic approach provides superior coverage of the search space, establishing a more favorable starting point for the subsequent optimization process.

\subsection{VM Load-Aware Redistribution Policy}
During the scheduling process, both WOA and SOA generate candidate solutions that map tasks to VMs. However, some candidate solutions may violate VM capacity constraints, necessitating corrective task reassignment. Traditional reassignment methods often rely on random allocation, disregarding current VM workloads. This frequently results in significant load imbalance, where some VMs become overloaded while others remain underutilized, thereby degrading overall system efficiency and resource utilization.

To address this issue, we propose a dynamic VM load-aware redistribution mechanism. This mechanism continuously monitors the workload of all VMs and maintains a real-time scheduling table that records accumulated task execution times, enabling precise tracking of resource consumption across the infrastructure. When constraint violation triggers a reassignment, VMs are sorted in ascending order based on their current load. Tasks from the infeasible solution are then systematically reassigned to the least-loaded VM that satisfies their resource requirements. This strategy actively promotes a balanced task distribution, minimizes resource idleness, and enhances overall scheduling performance and system throughput.

\subsection{Parallel Processing Acceleration}
Makespan is a critical performance metric in large-scale cloud task scheduling, as prolonged execution times directly reduce system throughput and degrade user experience. Traditional sequential scheduling algorithms often prove inefficient for high-dimensional, complex optimization problems, limiting their scalability in practical deployments.

To overcome this computational bottleneck, we develop a multithreaded parallel processing framework to significantly accelerate the optimization process. This framework strategically distributes the computationally intensive tasks of population evaluation and position updates across multiple concurrent threads. This allows for the simultaneous exploration of diverse regions within the search space, accelerating convergence without compromising solution quality. To ensure algorithmic robustness, a timeout mechanism (set to one minute) is implemented to prevent potential deadlocks during inter-thread synchronization \cite{12}.

By drastically reducing computational overhead and enabling more solution evaluations per unit time, our parallel framework effectively contributes to makespan minimization, providing a scalable and efficient scheduling solution for large-scale cloud environments.

\subsection{Introducing a variation operator}
Although the WOA demonstrates robust global exploration, it is prone to slow convergence and iterative stagnation, often yielding suboptimal solutions in later phases. To address these limitations, we integrate a mutation operator that injects controlled stochastic perturbations during the position update phase. This mechanism helps preserve population diversity, escape local optima, and enhance the overall global search efficiency of the optimization process.

The mutation mechanism operates with a predefined mutation probability $p_m$. Following each position update, the algorithm generates a random number $\mathbf{r} \sim U(0,1)$. If $\mathbf{r} < p_m$, the current solution undergoes mutation through uniform random sampling within the prescribed search boundaries $[lb, ub]$. This process can be formally expressed as:

\begin{equation}
X_{\text{mutated}} = lb + \mathbf{r} \cdot (ub - lb)
\end{equation}

where $\mathbf{r}$ is a random vector uniformly distributed in $[0, 1]^d$, with $d$ representing the problem dimensionality, and the multiplication is performed element-wise. By periodically introducing new candidate solutions through this mutation scheme, the algorithm effectively diversifies the population, explores uncharted regions of the search space, and accelerates convergence toward superior solutions \cite{13}.

\subsection{Maintaining the Pareto Solution Set}
This section formalizes the concept of Pareto optimality and details its integration into the proposed PHWSOA for multi-objective task scheduling. In multi-objective optimization, a solution is deemed Pareto-optimal (or non-dominated) if no other feasible solution exists that can improve at least one objective without worsening another. The set of all such Pareto-optimal solutions constitutes the Pareto frontier, which encapsulates the optimal trade-offs among the conflicting objectives \cite{14}.

As illustrated in the bi-objective scenario depicted in Figure 2, points lying on the lower-left convex boundary represent the Pareto-optimal set. Within this set, no solution is dominated by another in both objectives simultaneously.

\begin{figure}[htbp]
    \centering
    \includegraphics[width=0.7\linewidth]{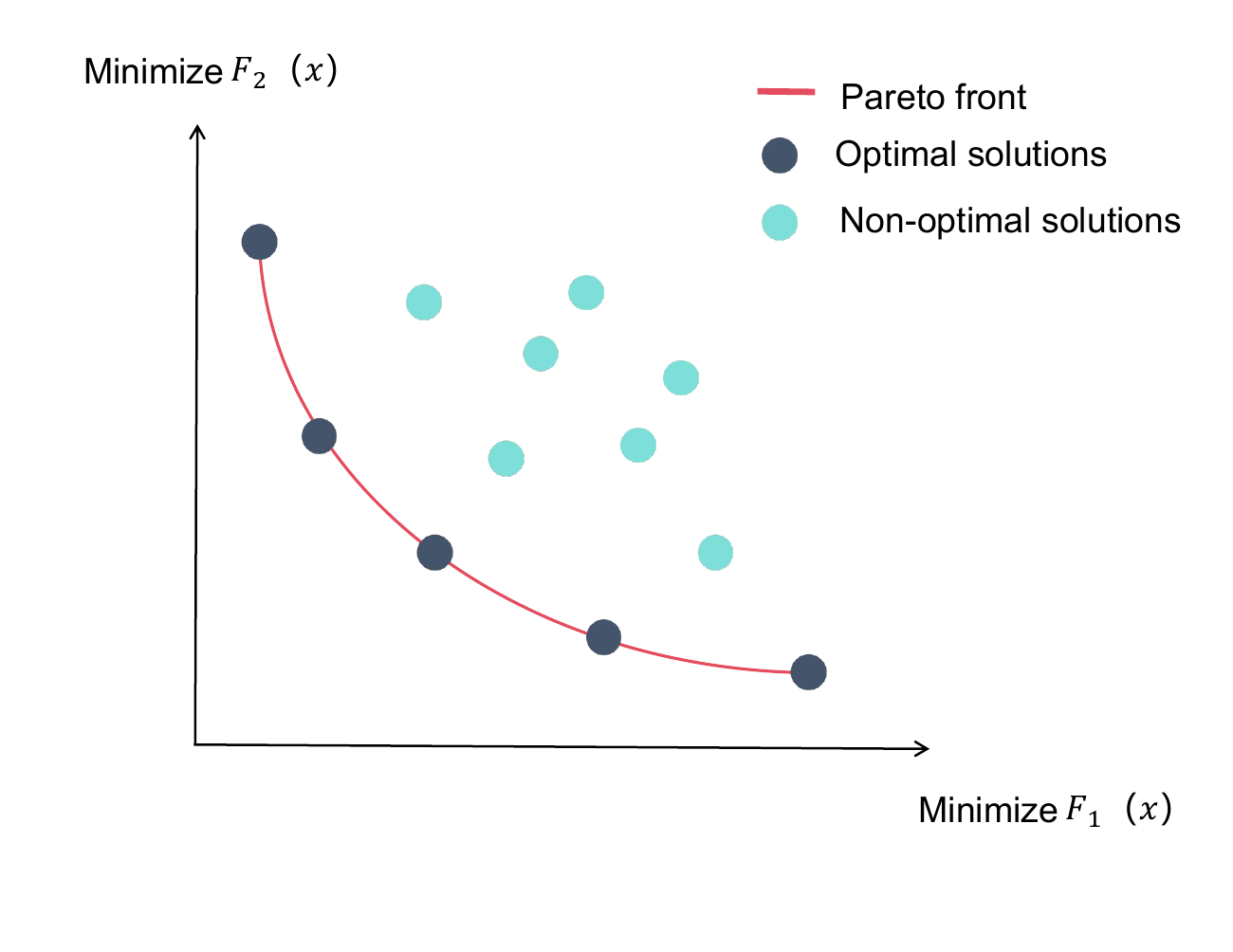}
    \caption{Example of dominated, non-dominated and pareto front solution set for a bi-criteria optimization problem. }
    \label{fig:enter-labe1}
\end{figure}

To embed Pareto optimality within the search process, PHWSOA maintains a dynamic external archive that stores non-dominated solutions throughout the optimization. In each iteration, candidate solutions generated by both whale and seagull agents are evaluated for Pareto dominance. Newly found non-dominated solutions are added to the archive, while any solutions in the archive that become dominated are removed. To control the archive size, a crowding-distance mechanism is employed; when the archive exceeds its predefined capacity, solutions with the smallest crowding distance (i.e., those in densely populated regions of the objective space) are pruned to maintain diversity.

After the algorithm converges, the solutions in the archive are ranked according to their Pareto dominance relations. The top 50 candidates—characterized by superior dominance ranks and diversity metrics—are selected to form the final approximated Pareto frontier, offering decision-makers a set of well-distributed optimal trade-off alternatives.

To select a single representative solution from this frontier for practical deployment, the three objectives—makespan, cost, and VM load balance—are first normalized to a common, dimensionless scale. The Mean Squared Deviation (MSD) is then computed for each Pareto-optimal solution as follows:

\begin{equation}
MSD = \frac{1}{3} \sum_{i=1}^{3} \left(\frac{f_i - f_i^{\min}}{f_i^{\max} - f_i^{\min}}\right)^2
\end{equation}

where $f_i$ represents the $i$-th objective value, and $f_i^{\min}$ and $f_i^{\max}$ denote the minimum and maximum values of the $i$-th objective across the Pareto set, respectively. The solution with the minimum MSD value is selected as the final scheduling decision, representing a balanced compromise across all optimization objectives.

\subsection{Pseudo-code and flowchart of the hybrid Pareto-based whale-seagull optimization algorithm}
This study implements the PHWSOA framework for multi-objective cloud task scheduling through the following sequential procedure:

(1) Initialization: Specify the number of tasks (nTasks) and virtual machines (nVMs). Initialize the whale and seagull populations using Halton sequences to ensure uniform coverage of the search space. Set algorithm parameters, including the maximum number of iterations, population size, and mutation probability.

(2) Pareto-based Elite Selection: Evaluate all candidate solutions against the multiple objectives. Identify the non-dominated set through pairwise Pareto dominance comparisons. These elite solutions are used to guide the population’s evolution while preserving diversity.

(3) Parallel Position Update: Employ parallel computing to concurrently update the positions of candidate solutions according to the movement mechanisms of WOA and SOA. Apply a mutation operator with probability $p_{\text{m}}$ to specific agents to maintain population diversity.

(4) Pareto Archive Maintenance: Update an external archive by adding newly discovered non-dominated solutions and removing any that become dominated. If the archive exceeds its capacity, a crowding-distance mechanism is triggered to prune solutions from the most crowded regions of the objective space.

(5) Constraint Handling and Load-aware Redistribution: Validate the feasibility of each solution by ensuring all task-to-VM assignments satisfy capacity constraints. If violations occur, dynamically redistribute tasks based on real-time VM workload to optimize resource utilization and balance load.

(6) Termination and Output: Execute the optimization for a predefined number of iterations. Upon termination, output the final approximated Pareto-optimal set. Record comprehensive performance metrics, including makespan, throughput, VM load distribution, and economic cost.

Algorithm 1 presents the complete pseudocode of PHWSOA, while Figure 3 illustrates its high-level execution workflow.

\begin{algorithm}
  \caption{Pareto-based Hybrid Whale-Seagull Optimization Algorithm}\label{algorithm}
  \SetKwInOut{Input}{input}\SetKwInOut{Output}{output}

  \Input{number of tasks (nTasks), number of VMs (nVMs). }
  \Output{optimalPos, makespan, throughput, loadCost, totalCost. }
  \BlankLine
  Initialize population using Halton sequence
  
  $popSize\leftarrow 50$
  
  $maxIter\leftarrow 100$
  
   \For{$t=1$ \KwTo $maxIter$}{
      Pool of threads (Whales, Seagulls)
      
      \For{each population}{
        \For{each task}{
            \eIf{$p < 0.5$}{
                \eIf{$|A| < 1$}{
                    $D_{leader}\leftarrow |C \times optimalWoa - curWoa|$\\
                    $newWoa \leftarrow | optimalWoa - A \times D|$
                }{
                    $D_{rand}\leftarrow |C \times randomPos - curWoa|$\\
                    $newWoa \leftarrow | randomPos - A \times D_{rand}|$
                }
            }{
                $D \leftarrow |optimalPos - curPos|$\\
                $newWoa \leftarrow D \times e^{bl} \times \cos(2\pi l) + optimalWoa$
            }
        }
        \For{each task}{
            $C_s \leftarrow A \times curSoa$\\
            $M_s \leftarrow B \times (optimalSoa - curSoa)$\\
            $D_s \leftarrow |C_s + M_s|$\\
            $newSoa \leftarrow D_s \times x \times y \times z + optimalSoa$\\
            \If{$newWoa > ub$ \textbf{or} $newWoa < lb$}{
                $newWoa \leftarrow vmSort$\;
            }
        }
      }
      executor.shutdown()\\
      calcPopulationFitness()\\
      archiveUpdate()
    }
\end{algorithm}

\begin{figure}[htbp]
    \centering
    \includegraphics[width=0.7\linewidth]{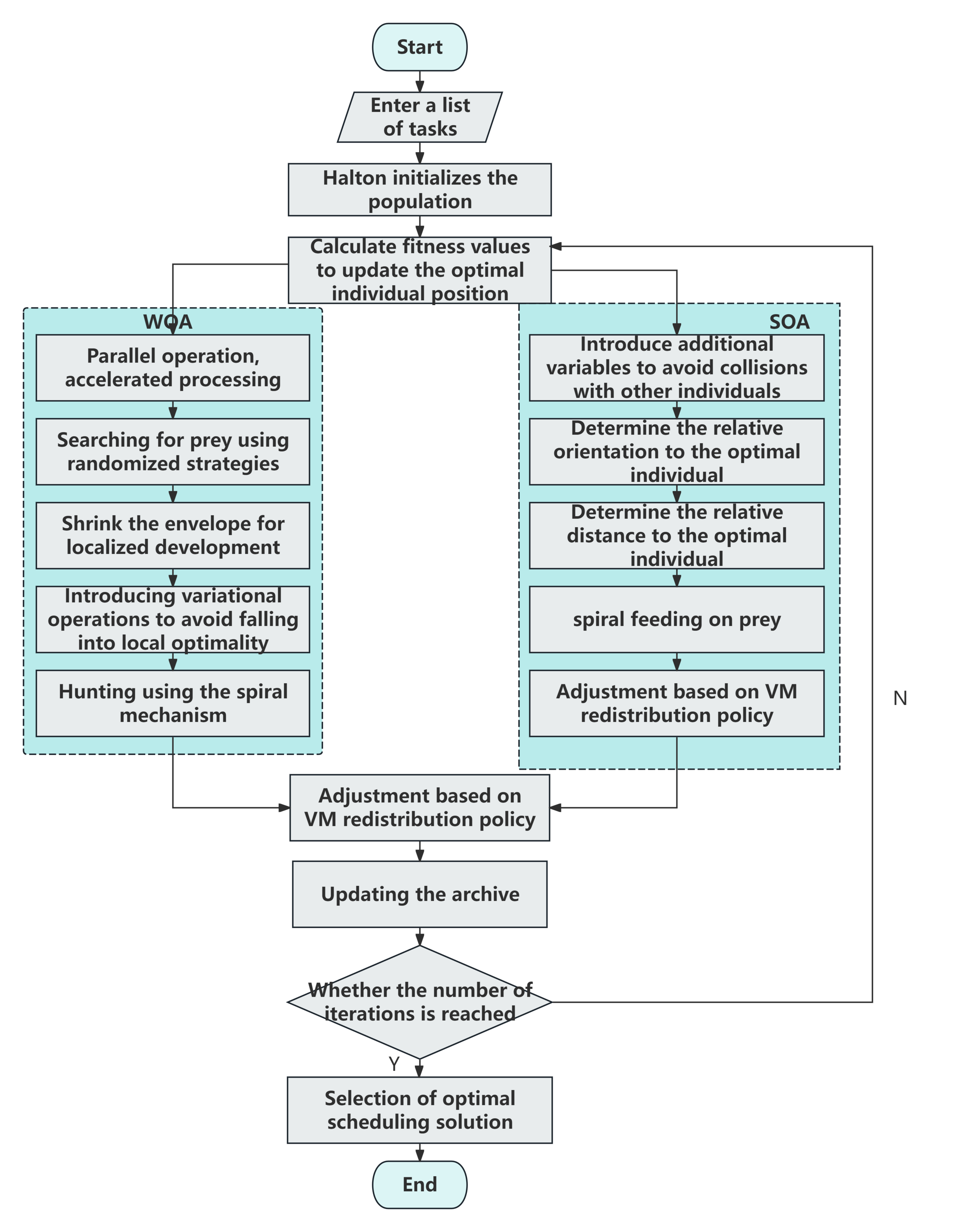}
    \caption{Flowchart of PHWSOA algorithm. }
    \label{fig:enter-labe1}
\end{figure}

\section{Simulation and Results}
This section presents the experimental evaluation of PHWSOA using the CloudSim simulation toolkit. All simulations were conducted on a computing platform equipped with an Intel Core i5-9300H processor (2.40 GHz), 8 GB of RAM, and a 512 GB SSD. The experiments employed real-world workload traces from the NASA-iPSC and HPC2N datasets. The performance of PHWSOA was compared against several established heuristic and meta-heuristic baseline algorithms: GA, the WOA, and the enhanced variants PEWOA and GCWOA. Four key scheduling metrics were assessed: makespan, system throughput, VM load balance, and total economic cost. The complete simulation parameters and environmental configurations are summarized in Table 1.
\begin{table}[htbp]
\centering
\caption{Configuration settings used for simulation. }
\begin{tabular}{ll}
\hline 
Parameters & Configuration Value  \\
\hline 
\# NASA iPSC & 50-500  \\
\# HPC2N & 500-5000 \\
\# Virtual Machines & 32-64 \\
Task Length & 15000-500000MI \\
\# Processing elements & 500-10000Mips \\
Memory capacity of Virtual Machines & 512MB \\
Storage capacity of Virtual Machines & 3072-10240MB \\
Bandwidth capacity of Virtual Machines & 1000Mbps\\
Name of Hypervisor & Xen \\
\hline 
\end{tabular}
\label{tab:simulation_config}
\end{table}

\subsection{Small-scale task}
Scheduling experiments were conducted using the NASA-iPSC dataset, with the number of tasks varying from 100 to 500 in increments of 100. Figure 4(a) illustrates that the makespan increases monotonically with the task count for all evaluated algorithms. PHWSOA consistently achieves the shortest makespan, exhibiting an average reduction of 62.7\% compared to the ensemble of baseline methods. This substantial improvement is attributed to its parallel processing framework and the synergistic integration of global exploration (from WOA) and local exploitation (from SOA).
Among the baseline algorithms, the enhanced WOA variants-PEWOA and GCWOA-show competitive performance, which can be traced to their advanced initialization strategies and adaptive parameter tuning. While the standard WOA underperforms relative to these enhanced versions, it still surpasses the traditional GA, underscoring the general efficacy of population-based metaheuristics in this domain.

Figure 4(b) presents the system throughput results. For smaller task volumes (e.g., 100 tasks), the performance advantage of PHWSOA is constrained by limited opportunities for parallel execution. As the task count increases, PHWSOA's throughput rises substantially, reaching approximately 80\% under the 500-task scenario. This demonstrates excellent scalability, stemming directly from its parallel execution framework and hybrid search strategy. In contrast, the throughput of PEWOA and GCWOA declines as the workload increases. Both the standard WOA and GA maintain consistently low throughput levels (around 20\%), due to their static search mechanisms and inherent lack of parallel processing capabilities.

The VM load balancing performance is shown in Figure 4(c). PHWSOA achieves a reduction in load variance of approximately 35\% compared to the baseline methods. This improvement is sustained as task volumes grow, enabled by the algorithm's real-time monitoring and adaptive task redistribution mechanism. Enhanced load balance directly contributes to greater system stability and resource utilization efficiency. Conversely, GA's static allocation strategy leads to poor load distribution and extended makespan, resulting in inferior overall performance.

Figure 4(d) summarizes the economic cost analysis. PHWSOA achieves an execution cost reduction of 13.7\% compared to the best-performing baseline. Its adaptive resource allocation and dynamic parameter tuning mechanisms effectively minimize operational overhead, ensuring that cost efficiency scales favorably with increasing task volumes. In comparison, GA incurs the highest operational costs, a consequence of its propensity for premature convergence and inefficient resource utilization, which collectively prolong execution time and amplify computational expenses.

\begin{figure}[htbp]
    \centering
    \includegraphics[width=1\linewidth]{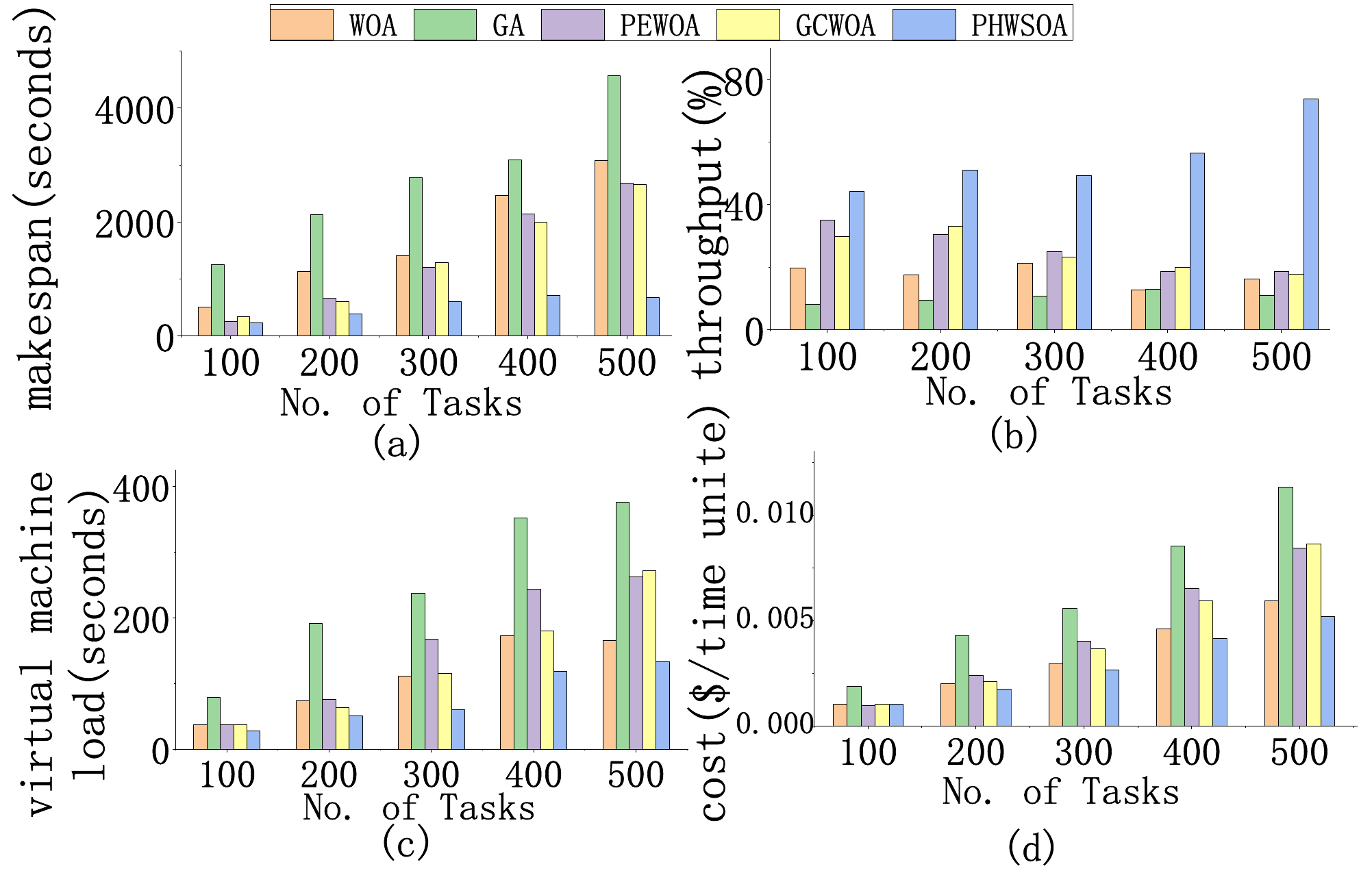}
    \caption{Tasks on NASA-iPSC instances. (a) makespan, (b) throughpu, (c) virtual machine load, (d) cost. }
    \label{fig:enter-labe1}
\end{figure}

\subsection{Large-scale task}
To rigorously evaluate algorithmic scalability, we conducted extensive scheduling experiments using the large-scale HPC2N dataset. The number of tasks was scaled from 1,000 to 5,000 in increments of 1000. Figure 5 presents a comparative performance analysis of the five algorithms across the four core metrics: makespan, throughput, load balancing efficiency, and economic cost.

PHWSOA demonstrates exceptional scalability, achieving an average reduction of 81.5\% in makespan, a 73.2\% improvement in throughput, a 35.8\% enhancement in load balancing efficiency, and a 33.2\% decrease in economic cost relative to the best-performing baseline. These gains substantially exceed those observed in the NASA-iPSC experiments, underscoring PHWSOA's superior aptitude for large-scale scheduling scenarios.

The performance superiority stems from several key algorithmic innovations: (1) the parallel processing framework efficiently manages the growth in computational complexity with problem scale; (2) the hybrid global-local search strategy preserves search efficacy in high-dimensional solution spaces; (3) the dynamic VM load redistribution mechanism guarantees efficient resource utilization under heavy workloads; and (4) the Pareto-based solution selection ensures balanced optimization across all competing objectives.

Notably, while the enhanced WOA variants (PEWOA and GCWOA) show moderate scalability, their performance degrades more rapidly than PHWSOA's when task counts exceed 2,000, revealing limitations in their adaptive mechanisms. The standard WOA and GA exhibit the poorest scalability, with key performance metrics deteriorating significantly at larger scales due to an inadequate exploration-exploitation balance and a lack of advanced load-management strategies.

In summary, these results confirm that PHWSOA effectively addresses the core challenges of large-scale cloud task scheduling. It maintains robust and stable performance across all evaluated metrics, whereas the comparative algorithms show pronounced degradation as the problem size increases.

\begin{figure}[H]
    \centering
    \includegraphics[width=1\linewidth]{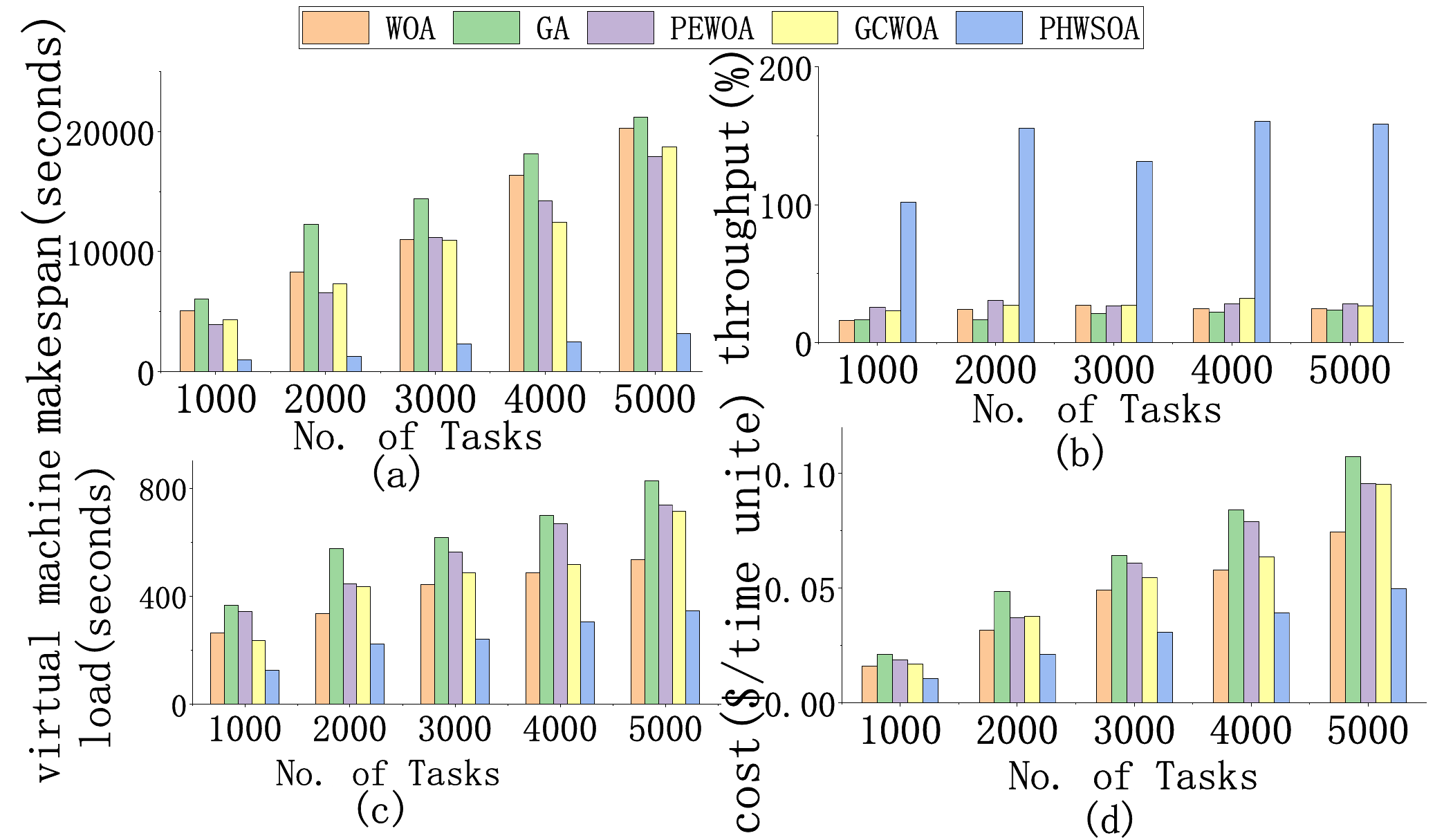}
    \caption{Tasks on HPC2N instances. (a) makespan, (b) throughpu, (c) virtual machine load, (d) cost. }
    \label{fig:enter-labe1}
\end{figure}

\section{Conclusion}
Cloud computing platforms suffer performance degradation in large-scale task scheduling caused by exponential task growth, leading to longer makespan, higher costs, and significant VM load imbalance. To address these challenges, we propose the PHWSOA, integrating WOA's global exploration with SOA's local exploitation. PHWSOA employs Pareto dominance to optimize makespan, VM load balancing, and economic cost simultaneously. It uses Halton sequence initialization to improve population diversity and introduces dynamic VM load redistribution and mutation operators to avoid premature convergence. A parallel processing framework further accelerates convergence and enhances computational efficiency. Experiments on CloudSim with NASA-iPSC (small-scale) and HPC2N (large-scale) workload traces show PHWSOA consistently outperforms baselines (WOA, GA, PEWOA, GCWOA). PHWSOA achieves up to 72.1\% makespan reduction, 63.8\% throughput increase, 36.8\% load variance decrease, and 23.5\% cost reduction, with statistically significant gains across all scenarios.

\bibliographystyle{ieeetr}
\bibliography{reference}

\end{document}